%% file: 2023_FeAR.tex
\documentclass{ecai}

\usepackage{graphicx}
\usepackage{amsmath}
\usepackage{booktabs}
\usepackage{longtable}
\usepackage{tabularx}
\usepackage{hyperref}
\hypersetup{
	colorlinks   = true, 
	urlcolor     = black, 
	linkcolor    = black, 
	citecolor   = black 
}

\usepackage{xparse}
\usepackage{comment}
\usepackage{algorithm}
\usepackage{algpseudocode}
\usepackage[capitalise]{cleveref}
\usepackage{subcaption}
\usepackage[dvipsnames]{xcolor}

\usepackage{import}
\import{./}{agdefs.tex}

\begin{document}
	
\begin{frontmatter}

\title{Feasible Action-Space Reduction as a Metric of Causal Responsibility in Multi-Agent Spatial Interactions}


\author[A]{\fnms{Ashwin}~\snm{George}\thanks{Corresponding Author. Email: A.George@tudelft.nl}}
\author[A]{\fnms{Luciano}~\snm{Cavalcante Siebert}}
\author[A]{\fnms{David}~\snm{Abbink}}
\author[A]{\fnms{Arkady}~\snm{Zgonnikov}}

\address[A]{Delft University of Technology}

\begin{abstract}
Modelling causal responsibility in multi-agent spatial interactions is crucial for safety and efficiency of interactions of humans with autonomous agents. However, current formal metrics and models of responsibility either lack grounding in ethical and philosophical concepts of responsibility, or cannot be applied to spatial interactions. In this work we propose a metric of causal responsibility which is tailored to multi-agent spatial interactions, for instance interactions in traffic. In such interactions, a given agent can, by reducing another agent's feasible action space, influence the latter. Therefore, we propose feasible action space reduction (FeAR) as a metric of causal responsibility among agents. Specifically, we look at \mbox{ex-post} causal responsibility for simultaneous actions. We propose the use of \textit{Moves de Rigueur} (MdR) --- a consistent set of prescribed actions for agents --- to model the effect of norms on responsibility allocation. We apply the metric in a grid world simulation for spatial interactions and show how the actions, contexts, and norms affect the causal responsibility ascribed to agents. Finally, we demonstrate the application of this metric in complex multi-agent interactions. We argue that the FeAR metric is a step towards an interdisciplinary framework for quantifying responsibility that is needed to ensure safety and meaningful human control in human-AI systems.
\end{abstract}

\end{frontmatter}

\section{Introduction}
\label{Sec:Introduction}

The rapid development of AI methods have sparked debates on the ethics of AI \cite{calvertHumanCentricFramework2020, cavalcantesiebertMeaningfulHumanControl2022,flemischDynamicBalanceHumans2012, santonidesioEuropeanCommissionReport2021}. Ethical principles like responsibility, trust, and fairness need to be operationalised for making AI systems sensitive to human values \cite{europeancommission.directorategeneralforcommunicationsnetworkscontentandtechnology.EthicsGuidelinesTrustworthy2019}. In particular, understanding responsibility of agents is crucial for designing autonomous systems and distributing rewards and punishments during their operation in an appropriate way. For example, consider an automated vehicle (AV) that brakes abruptly to prevent hitting a child that walked on to the road. Who would be responsible if the AV ends up in a rear-end collision? How should the AV be designed to prevent situations when no agent is actively responsible for preventing negative outcomes?~\cite{santonidesioMeaningfulHumanControl2018} Understanding the responsibility of agents is therefore crucial for the users, designers and regulatory bodies of such an automation \cite{beckersDriversPartiallyAutomated2022, papadimitriouCommonEthicalSafe2022, santonidesioEuropeanCommissionReport2021}.   

On the practical side, researchers working on design and engineering of automated driving systems have been trying to incorporate the notion of responsibility in motion planning frameworks for such systems. For example, the Responsibility-Sensitive Safety (RSS) approach formalises an interpretation of "Duty of Care" from tort law and promises verifiable safety when all agents are following RSS \cite{shalev-shwartzFormalModelSafe2018, shalev-shwartzVisionZeroProvable2019}. This approach provides the "proper responses" for all the agents and those deviating from the "proper response" would be considered responsible if something goes wrong. However, for mixed traffic (in which both AVs and human-driven vehicles share the road), not all agents might know the "proper response" or be capable of executing it. Hence, this approach is not ideal for mixed-traffic with humans having variable behaviours. Another approach is to include an extra term in the cost function of trajectory planning algorithms of AVs to account for responsibility. This term can incorporate an interpretation of responsibility based on traffic rules and physical constraints~\cite{geisslingerEthicalTrajectoryPlanning2023}, or be learned from driving data~\cite{cosnerLearningResponsibilityAllocations2023}. However, the simplified notions of "responsibility" captured by these approaches are not grounded in the established definitions of responsibility from ethics and philosophy. 

On the more fundamental side, metrics and models of responsibility have been developed that aim to quantify different flavours of responsibility, e.g., moral responsibility, legal liability, blameworthiness, praiseworthiness, role responsibility, and causal responsibility \cite{vincentStructuredTaxonomyResponsibility2011}. Of these, causal responsibility --- the causal importance of an agent’s action for the implementation of an event \cite{englTheoryCausalResponsibility2018} --- is a prerequisite for other forms of responsibility, including moral responsibility and blame \cite{vandepoelEthicsTechnologyEngineering2011, vincentStructuredTaxonomyResponsibility2011}, and therefore received most attention in terms of existing metrics and models. Models of causal responsibility traditionally involve counterfactual and probabilistic reasoning about an agent's action being pivotal for an event, \cite{bartlingShiftingBlameDelegation2012, chocklerResponsibilityBlameStructuralModel2004, englTheoryCausalResponsibility2018, triantafyllouActualCausalityResponsibility2022}. Collective causal responsibility of groups of agents in an interaction has also been modelled by exploring how groups of agents by concerted actions can cause or prevent an event \cite{yazdanpanahDistantGroupResponsibility2016, yazdanpanahDifferentFormsResponsibility2021, yazdanpanahApplyingStrategicReasoning2021}. Specifically in the context of human-AI interactions, comparative human causal responsibility has been modelled using information theory to quantify the influence of human agents on an outcome~\cite{douerResponsibilityQuantificationModel2020, douerJudgingOneOwn2022, douerTheoreticalMeasuredSubjective2021}. These more fundamental metrics and models are suitable for relatively abstract scenarios where agents can choose not to act at a given instant. However, in spatial interactions like traffic, all the agents are physically embedded in the scene and they cannot choose not to act --- not moving or not stopping would also have an impact on the safety of the interaction. This makes it difficult to apply the existing work on quantifying causal responsibility to spatial interactions of embodied agents.  

\noindent
\textbf{Contribution. }
To bridge the gap between the existing practical approaches to quantifying responsibility in traffic and the theoretically-grounded metrics of responsibility, here we propose a metric for causal responsibility designed specifically for spatial interactions. We assume perfect information conditions and simultaneous actions to propose a metric for ex-post causal responsibility for one instant of interaction. The metric is based on the principle that an agent restricting the feasible action space of another agent should be held responsible for the state and safety of the latter. To make the metric sensitive to norms, we propose the use of \textit{Moves de Rigueur} --- a consistent set of default actions that are prescribed for all the agents. 

\noindent
\textbf{Outline. }
In \cref{Sec:GridWorld} we describe the nature of spatial interactions and how we model these in a grid world simulation. \cref{Sec:FeARMetric} introduces the Feasible Action-Space Reduction (FeAR) metric and the notion of \textit{Move de Rigueur} (MdR). The sensitivity of the FeAR metric to actions and norms, and a potential application of the metric to multi-agent interactions are demonstrated with case studies in \cref{Sec:CaseStudies}. These are followed by the discussion and conclusions (\cref{Sec:Discussion}).

\section{Grid World for Spatial Interaction}
\label{Sec:GridWorld}

\begin{figure}[t]
\centering
	\begin{subfigure}[b]{0.4\textwidth}
		\centering
		\includegraphics[width=\linewidth]{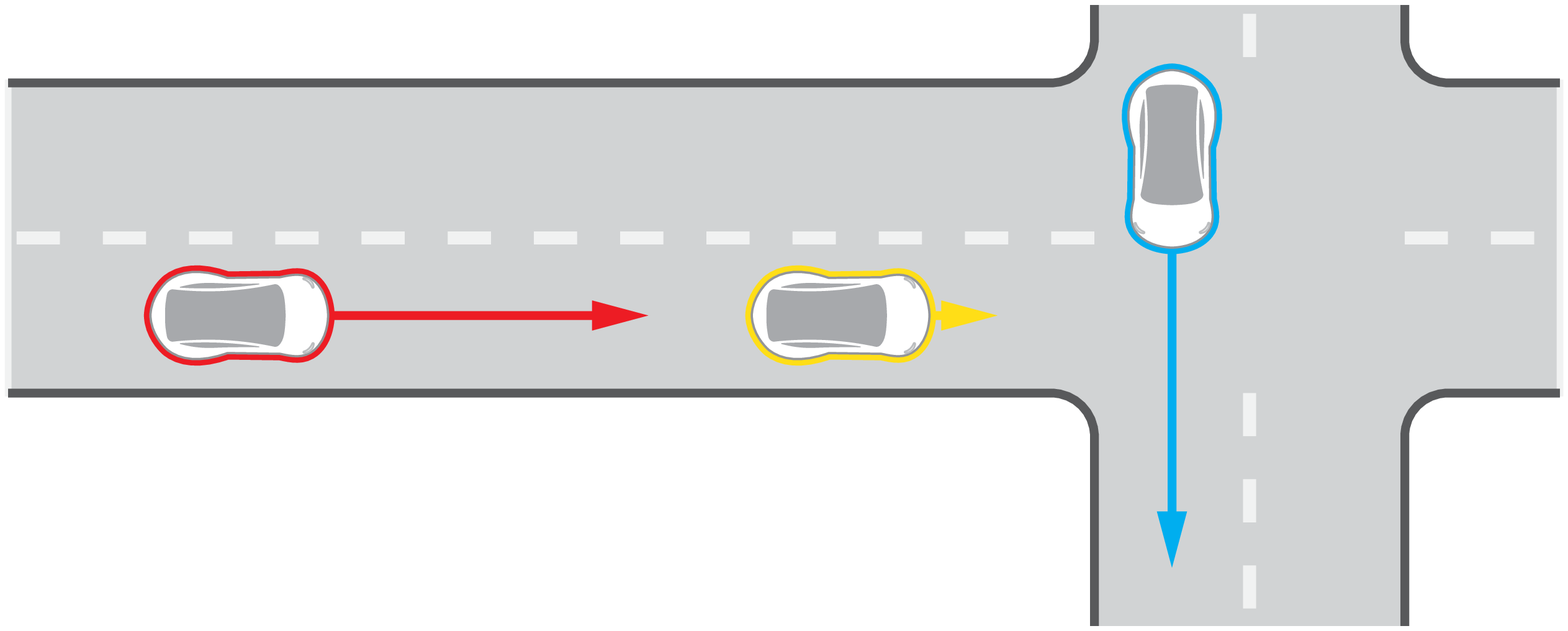}
		\caption[Real World]{Real world interaction}
		\label{fig:GWorld - R}
	\end{subfigure}
	\hfil
	\begin{subfigure}[b]{0.4\textwidth}
		\centering
		\includegraphics[width=\linewidth]{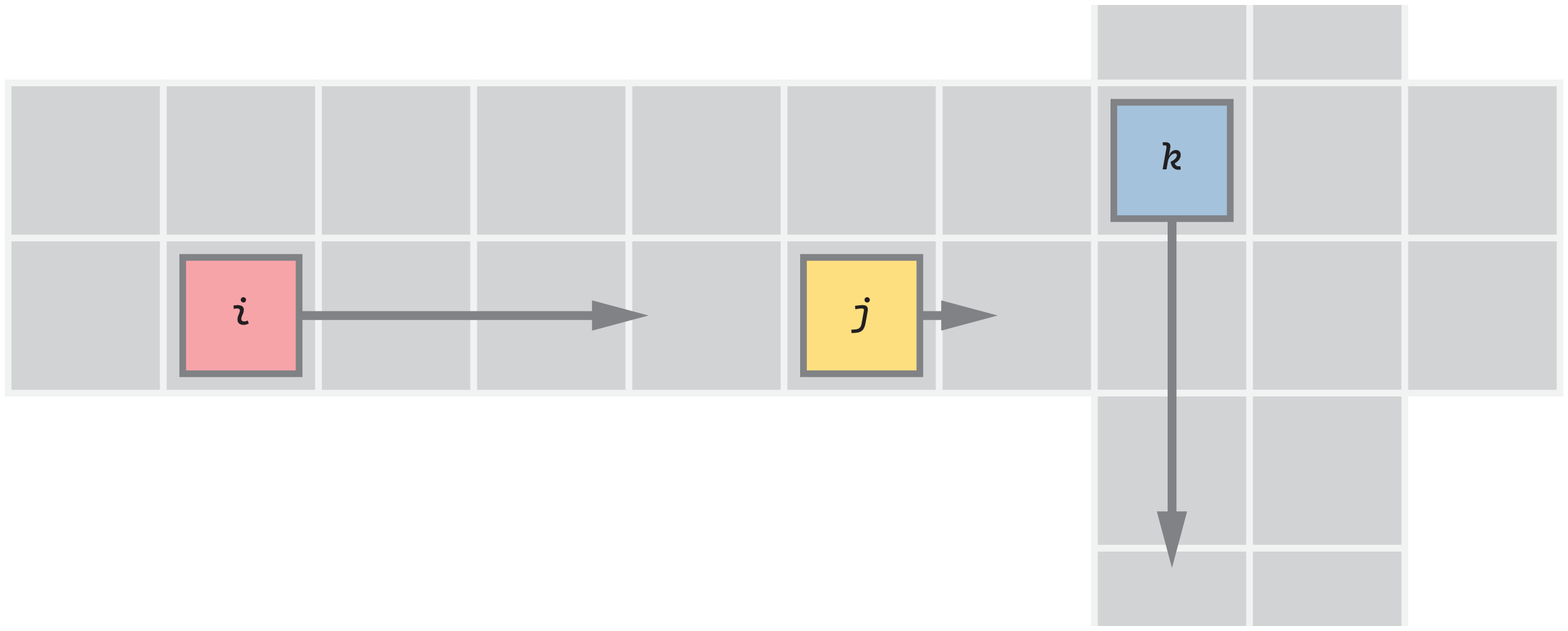}
		\caption[Grid World]{Grid world representation}
		\label{fig:GWorld - G}
	\end{subfigure}
	\label{fig:GandRworlds}
	\caption[The grid-world representation of a real-world interaction at an intersection.]{The grid-world representation of a real-world interaction at an intersection. The yellow car is slowing down at the intersection because the blue car is aggressively entering the intersection. This in turn affects the red car which has to slow down so as not to hit the yellow car. The red, yellow and blue cars are represented by agents $i$, $j$ and $k$ respectively in the grid world.}
\end{figure}

To define the metric, we need a model environment that captures the essence of spatial interactions. In spatial interactions the assertive moves of one agent can restrict the action space of another agent and the latter may be forced to move out of the way. Consider two cars approaching an intersection (yellow and blue cars in \cref{fig:GWorld - R}). If the blue car accelerates to cross first, it usually forces the yellow car to decelerate. On the hand, if the blue vehicle decelerates, it gives the yellow car more freedom to decide whether to accelerate or decelerate. Thus when the blue car accelerates, it is acting assertively because it has more influence on the state of the yellow car than when it decelerates. Now consider a third car (red) which is following the yellow car. If the yellow car slows down, the red car is also forced to slow down. Thus the assertiveness of the blue car can have an impact on the red car too. To model these interactions, it is necessary to consider the speed of the cars and the collisions between them in multi-agent settings.

We use grid world simulations to model multi-agent spatial interactions like the one described above. For example, the real-world scenario depicted in \cref{fig:GWorld - R} would be represented by the grid world instance shown in \cref{fig:GWorld - G}. We only consider the position and speed of the agents --- we assume that agents move with constant velocity in each instance. The actions chosen by the agents would correspond to their speed. An action would always be in one direction, and the speed would correspond to the number of cells an agent would move through during an action. For our cases, we assume that all agents are identical and have the same choice of actions. 

The maximum speed corresponds to actions that move through four cells. The actions are represented using the first letter of the direction followed by the number of cells it moves through - S0 for stay, R1 for one cell to the right, L1 for one cell to the left, U1 for one cell up, D1 for one cell down, etc. So each agent has 17 possible actions - S0, R1, R2, R3, R4, L1, L2, L3, L4, U1, U2, U3, U4, D1, D2, D3 and D4. For each instant, collision checks are done to see if any agents are colliding as they move. When a collision is detected, the agents that collide would stop at the last safe cell they occupied. Agents would be embedded in a map with valid locations. Actions of agents that try to push them outside the valid locations would also be treated as causing collisions. Details of the collision-check and step-update algorithms can be found in the supplementary materials\footnotemark. 

\section{The $\textbf{{FeAR}}$ Metric}
\label{Sec:FeARMetric}

\subsection{Preliminaries}
We consider a grid world with a state-space $\StateSpace$ where, $\aState \in \StateSpace$ represents the map of the grid world and the locations of agents. $\Agents$ represents the set of $k$ agents $\{1,..,k\}$. $\ActionSpace{i}$ is the action space of agent $i$. The joint action space of the all the agents is $\JointActionSpace = \times_{i=1}^{k} \ActionSpace{i}$. The action chosen by agent $i$ is $\action{i}$ and the joint action of all the agents is $\jointAction  = \times_{i=1}^{k} \action{i}$. Each agent in prescribed a \textit{Move de Rigueur} (MdR) $\mdr{i}(\aState)$ for a given state $\aState$. The joint MdR for all the agents is $\JointMdR(\aState)  = \times_{i=1}^{k}  \mdr{i}(\aState)$. We only consider one state per case, so for brevity from here on, we will refer to $\mdr{i}(\aState)$ as $\mdr{i}$ and $\JointMdR(\aState)$ as $\JointMdR$. We assume that the states and actions are finite and discrete. We also assume that each agent chooses one action per instance. 

For a given state $\aState$ and joint action $\jointAction$, the function $\validCount{\aState}{\jointAction}{j}$ gives the count of feasible moves available for agent $j$\footnotemark[\value{footnote}]\footnotetext{Detailed explanation of the functions and algorithms can be found in the supplementary materials: \url{https://osf.io/5s9m2}.}. An intervention of replacing $\action{i}$ (the move of agent $i$) with another action $a'_{i}$ is denoted as $ \jointAction_{i} \leftarrow a'_{i}$. Thus, the intervention of replacing $\action{i}$ with $\mdr{i}$ (the MdR of agent $i$) would then be $ \jointAction_{i} \leftarrow \mdr{i}$. $\nijMdR$ gives the number of feasible actions of agent $j$ when agent $j$ would have chosen the MdR $\mdr{i}$. $\niiMdR$ gives the number of feasible moves available to agent $i$ when all other agents $\neg i$, choose their MdRs $\mdr{\neg i}$. To map the value of the metric to the range $[-1, 1]$, the function $Z()$ was used.  
\begin{equation}
	Z(x) = 
	\begin{cases}
		1,& \text{if } x \geq 1\\
		x,& \text{if } -1 < x < 1\\
		-1,& \text{if } x \leq -1\\
	\end{cases}
\end{equation}

\subsection{Feasible Action-Space Reduction}

This work proposes a metric based on the idea that one agent taking actions that restrict the feasible actions of another agent must be held responsible (at least to some extent) for the state of the latter agent. Intuitively, in spatial interactions, an agent's action that leaves another agent with zero feasible moves, makes the latter unsafe in all possible contingencies, and thus should bear total responsibility for the state of the second. Also, we intuit that when the actions of the former do not change the feasible action space of the latter, the former has no influence on the state of the latter.  In the intermediate case, a greater reduction in the feasible action space of the latter should reflect in greater influence of the former on the latter. In this paper, when the action of an agent $i$ reduces the feasible action space of an agent $j$, we say that $i$ is \textit{assertive}  towards $j$. Similarly, when the action of agent $i$ increases the feasible action space of $j$, we say that $i$ is \textit{courteous} towards $j$. Based on these intuitions, we propose a metric based on feasible action space reduction to capture the influence of one agent on another.

Responsibility ascribed to agents is highly dependent on contexts and norms \cite{dastaniResponsibilityAISystems2023, lohResponsibilityRobotEthics2019, vincentStructuredTaxonomyResponsibility2011}. Agents in particular situations might experience a reduction in feasible actions that are not caused by the actions of other agents. For example, an agent in a dead-end might have a very small feasible action space. Therefore to capture the effect of contexts, we formulate the metric as the ratio of feasible action space reduction in comparison to a nominal case. To define the nominal case, we utilise the idea of \textit{Moves de Rigueur} (MdR) --- a consistent set of default actions that are prescribed to agents. Thus the \textit{\textbf{Fe}asible \textbf{A}ction-space \textbf{R}eduction} (FeAR) metric is defined based on the ratio by which the action of an agent $i$ reduces the feasible action space of another agent~$j$ ($\nij$) in relation to the feasible action space of the latter when $i$ would have chosen the MdR ($\nijMdR$) as: 

\begin{equation}
	\FeAR_{i,j} =
	\begin{cases}
		\clip{\frac{\nijMdR - \nij}{\nijMdR + \epsilon}}, & \text{if } i \neq j,\\
		\\
		\clip{\frac{\nii}{\niiMdR + \epsilon}}, & \text{if } i = j.\\
	\end{cases}
 \label{Eq:FeAR}
\end{equation}

 $\epsilon$ is added to the denominator to ensure that the metric can still be defined when $\nijMdR$ or $\niiMdR$ are zero ($0<\epsilon\ll1$).

Consider the FeAR ($\FeAR'_{i,j}$) for an alternative action $a'_{i}$ instead of $\action{i}$ for agent $i$. If $\FeAR'_{i,j} > \FeAR_{i,j}$, then $a'_{i}$ is more assertive towards $j$ than action $\action{i}$. If $\FeAR'_{i,j} < \FeAR_{i,j}$, then $a'_{i}$ is more courteous towards $j$ than action $\action{i}$. When $\FeAR_{i,j} = 1$, it means that $i$'s action reduces the feasible action space of $j$ to an empty set and thus $i$ should be ascribed complete responsibility for the state of $j$.

\subsection{Move de Rigueur}
The \textit{Move de Rigueur} $\mdr{i}$ is used to provide a baseline condition with which an agent's action is compared. In counterfactual analysis for causal responsibility, the pivotality of an action for an outcome is evaluated based on whether the outcome would have happened if the agent had not chosen that action \cite{halpernModificationHalpernpearlDefinition2015,englTheoryCausalResponsibility2018}. In a spatial interactions, all agents have to choose an action --- even if it is not to move or not to stop. We propose MdR as a means of relaxation of the counterfactual condition of not choosing an action.

In our definition, \textit{Moves de Rigueur} are a consistent set of  default actions that are prescribed to all the agents in an interaction. The prescribed MdRs have to be consistent in the sense that if all agents follow their MdRs, there should not be any collisions. The FeAR analysis is dependent on the assumption that all agents know their MdRs. We consider MdR as a heteronomous mandate \cite{graciaManyFacesAutonomy2012} which, if followed by the agents, must yield zero responsibility ascribed to these agents. Consequently, as per the definition \cref{Eq:FeAR}, $\FeAR_{i,j} (i \neq j)$ when $i$ chooses $\mdr{i}$ is zero.


$\FeAR_{i,i}$ indicates how much of the feasible action space of $i$ when all the other agents $\neg i$ had chosen their MdRs remains under the chosen joint action $\jointAction$. So, it is better to think of $\FeAR_{i,i}$ as the \textit{\textbf{Fe}asible \textbf{A}ction-space \textbf{R}emaining} for agent $i$. When $\FeAR_{i,i} = 1$, agent $i$ has at least as many feasible actions as it would have had when all the others choose their MdR, i.e, $i$ has as much freedom as the MdRs can afford.

The greater the FeAR caused by $i$ on $j$, the greater the influence of $i$ on $j$. $\FeAR_{i,j} > 0$ indicates that, agent $i$ choosing the current action, offers fewer actions to $j$, than when $i$ would have chosen the MdR i.e., $i$'s action is more assertive than the MdR. Similarly X. 

\section{Case Studies}
\label{Sec:CaseStudies}
In this section we use a series of simple case studies\footnote{The simulations were run on CPU (Intel i7) in a Dell Latitude 7320 laptop with 16 GB of RAM. The source code for reproducing all case studies can be found at \url{https://github.com/DAI-Lab-HERALD/FeAR}.} to demonstrate how the FeAR value is dependent on 1) the action chosen by the agent, 2) on the \textit{Moves de Rigueur} (MdR) and 3) on the actions of other agents. The last case study illustrates the use of the metric in multi-agent scenarios with large number of possible interaction outcomes.

\subsection{Two agents on a lane: the effect of agents' actions on FeAR}

\begin{figure*}[th]
	\centering
	
	\begin{subfigure}{0.4\linewidth}
		\centering
		\caption*{Grid world instance - state and action selection}
	\end{subfigure}
	\hfill
	\begin{subfigure}{0.2\linewidth}
		\centering
		\caption*{For actor's MdR $\mdr{i}$}
	\end{subfigure}
	\hfill
	\begin{subfigure}{0.2\linewidth}
		\centering
		\caption*{For actor's move $\action{i}$}
	\end{subfigure}
	\hfill
	\begin{subfigure}{0.12\linewidth}
		\centering
		\caption*{FeAR values}
	\end{subfigure}
\hrule
\hrule
\vspace{0.1cm}

	\begin{subfigure}{0.4\linewidth}
		\centering
		\includegraphics[width=\linewidth]{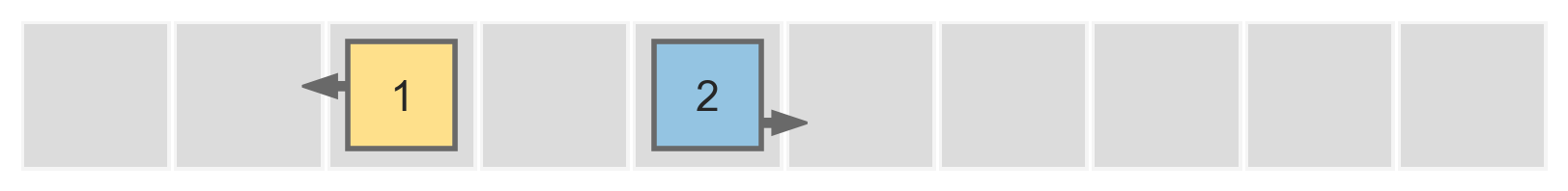}
		\caption{}
		\label{fig:2Agents-L1-R1}
	\end{subfigure}
	\hfill
	\begin{subfigure}{0.2\linewidth}
		\centering
		\includegraphics[width=\linewidth]{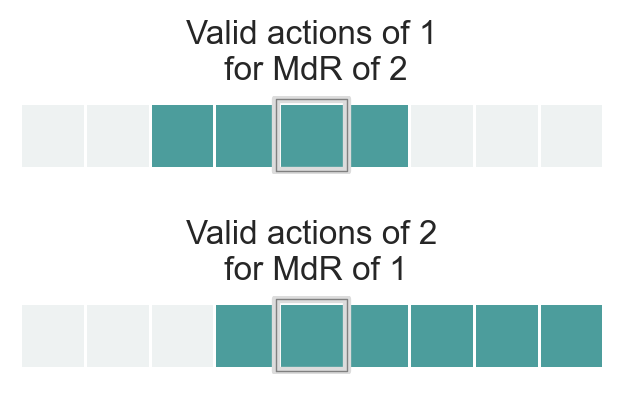}
	\end{subfigure}
	\hfill
	\begin{subfigure}{0.2\linewidth}
		\centering
		\includegraphics[width=\linewidth]{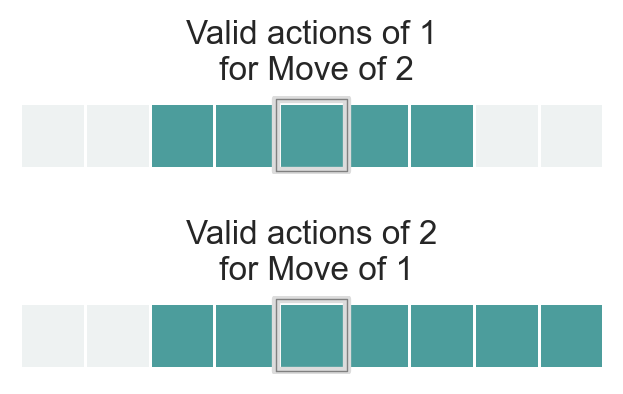}
	\end{subfigure}
	\hfill
	\begin{subfigure}{0.12\linewidth}
		\centering
		\includegraphics[width=\linewidth]{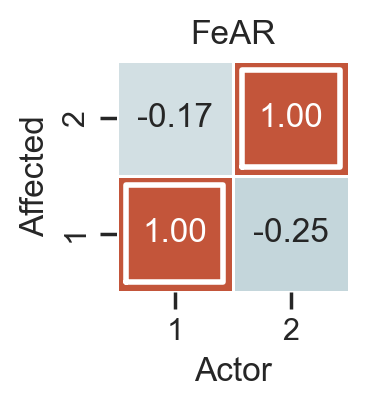}
	\end{subfigure}
\hrule
\vspace{0.01cm}

	\begin{subfigure}{0.4\linewidth}
		\centering
		\includegraphics[width=\linewidth]{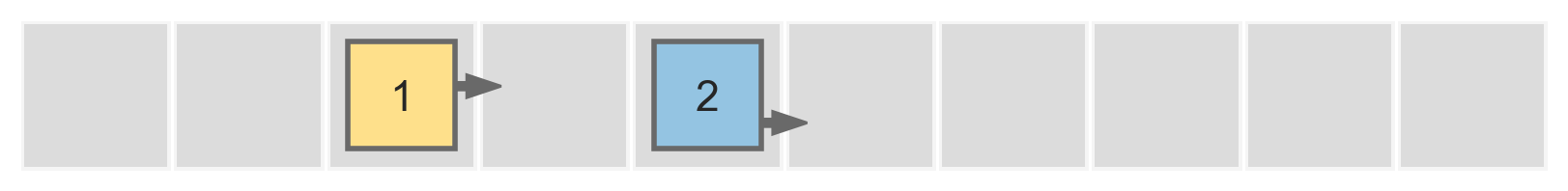}
		\caption{}
		\label{fig:2Agents-R1-R1}
	\end{subfigure}
	\hfill
	\begin{subfigure}{0.2\linewidth}
		\centering
		\includegraphics[width=\linewidth]{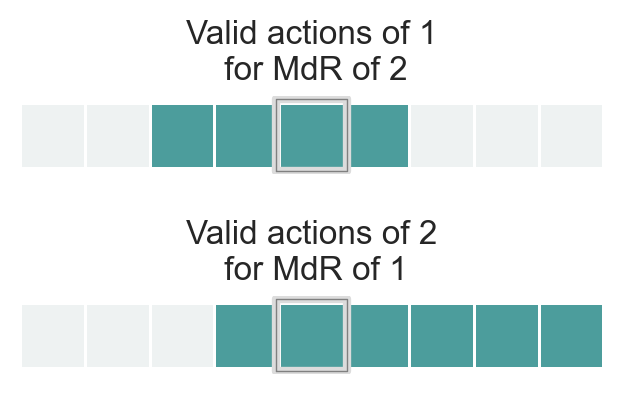}
	\end{subfigure}
	\hfill
	\begin{subfigure}{0.2\linewidth}
		\centering
		\includegraphics[width=\linewidth]{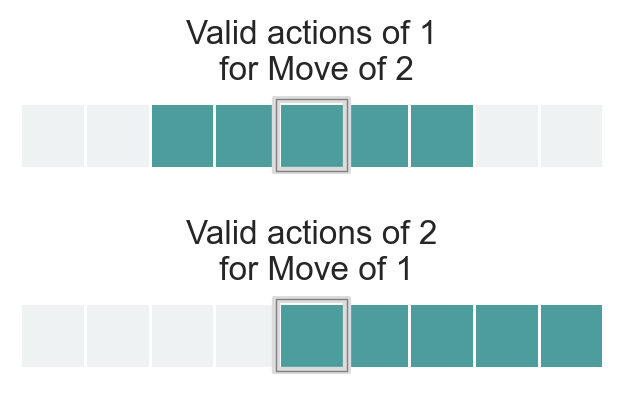}
	\end{subfigure}
	\hfill
	\begin{subfigure}{0.12\linewidth}
		\centering
		\includegraphics[width=\linewidth]{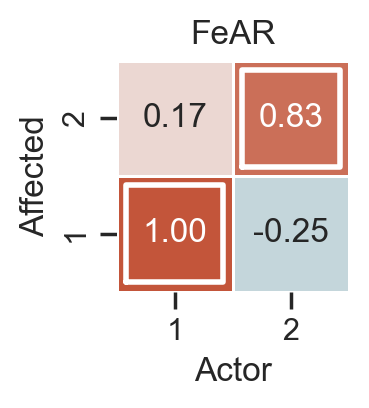}
	\end{subfigure}
\hrule
\vspace{0.01cm}

	\begin{subfigure}{0.4\linewidth}
		\centering
		\includegraphics[width=\linewidth]{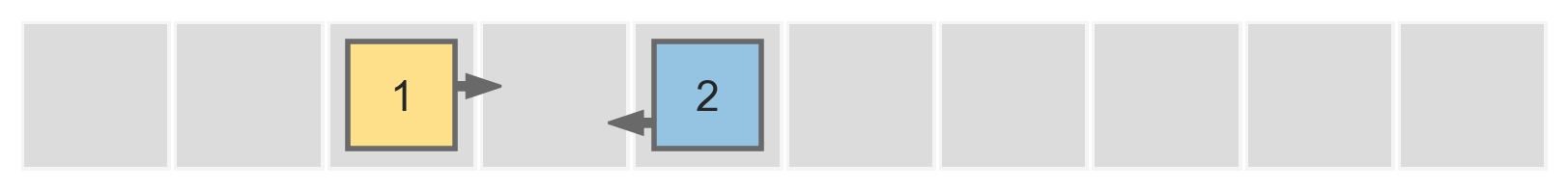}
		\caption{}
		\label{fig:2Agents-R1-L1}
	\end{subfigure}
	\hfill
	\begin{subfigure}{0.2\linewidth}
		\centering
		\includegraphics[width=\linewidth]{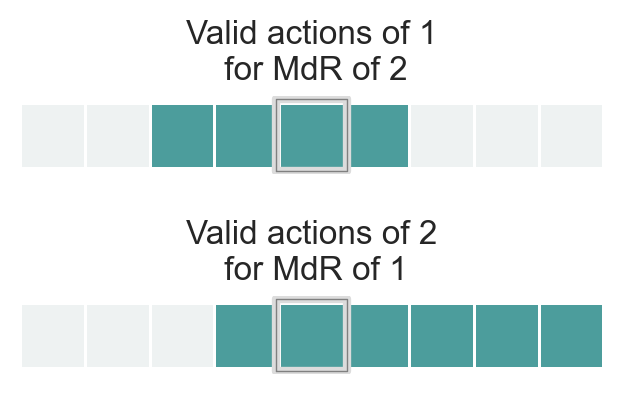}
	\end{subfigure}
	\hfill
	\begin{subfigure}{0.2\linewidth}
		\centering
		\includegraphics[width=\linewidth]{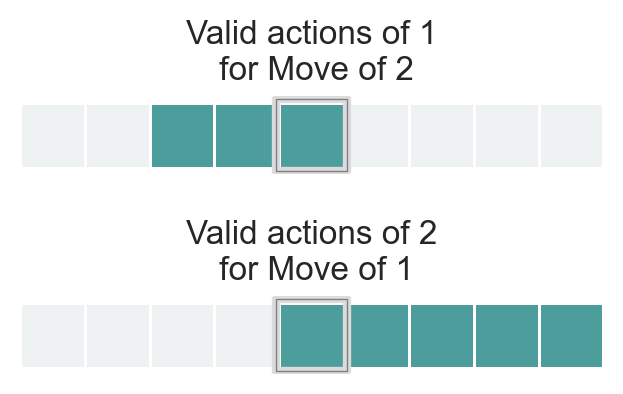}
	\end{subfigure}
	\hfill
	\begin{subfigure}{0.12\linewidth}
		\centering
		\includegraphics[width=\linewidth]{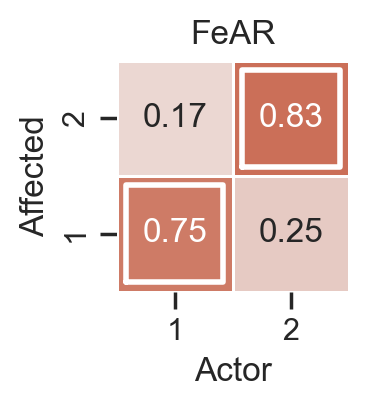}
	\end{subfigure}
\hrule
\vspace{0.01cm}
	\caption[Two agents on a lane]{\textbf{Case Study 1} explores the effect of an agent's action on the FeAR values for Mdr $\JointMdR = \text{S0-S0}$. The actions selected by the agents for each instance are depicted in the first column. The second and third columns show the feasible actions available to the agents when under the other agents'  MdR $\mdr{i}$  --- $\nijMdR$ --- or their chosen action $\action{i}$ --- $\nij$. The move $\text{S0}$ is depicted in the center with a grey rectangle around it with other left and right moves on either side. The feasible moves are shown in green. The final column states the FeAR values that are calculated based on the number of feasible actions. $\FeAR_{i,j}$ for $i \neq j$, represents the Feasible Action-Space Reduction of $j$ by the action of $j$. $\FeAR_{i,i}$ represent Feasible Action-Space Remaining for agent $i$ left by the actions of all other agents. Since the definition of $FeAR_{i,i}$ is slightly different, they are shown in the plots with an extra white rectangle around them. For $i \neq j$, if $\nij> \nijMdR$, then $\FeAR_{i,j} < 0$ and vice versa.} 
	\label{fig:2Agents}

\end{figure*}

The first case study demonstrates how the actions chosen by the agents affect the FeAR values. The location of agents 1 and 2 are asymmetric in the sense that 1 is more constrained than 2.  In \cref{fig:2Agents} we consider three instances where (a) both agents move one step away from each other, (b) both agents move one step to the right, and (c) both agents move a step closer to each other. In (a) and (b) there are no collisions while in (c) both agents collide. 

The MdR for the calculation is taken to be $\JointMdR=\text{S0-S0}$. When agent 1 chooses the MdR (S0), agent 2 has $\nijMdR= 6$ feasible moves (L1, S0, R1, R2, R3 and R4). When 2 chooses the MdR (S0), 1 has $\nijMdR= 4$ feasible moves (L2, L1, S0 and R1).

In Instance (a) 1 chooses L1 which leaves 2 with $\nij= 7$ feasible moves (L2, L1, S0, R1, R2, R3 and R4). So $\FeAR_{1,2} = \frac{6-7}{6} = -0.17$ \footnote{Z() only makes an impact when the magnitude of the ratios exceed 1. The $\epsilon$ in the denominator is only relevant in cases where the denominator is zero. For brevity, these are excluded in the explanations when unnecessary.}. And when 2 chooses R1, it increases the feasible actions to $\nij =5$ (L2, L1, S0, R1 and R2). Thus, $\FeAR_{2,1} = \frac{4-5}{4}= -0.25$. So both agents have negative FeAR and are courteous to each other. Since 1 has more feasible moves ($\nij = 5$) than when the other agent chooses the MdR ($\nijMdR = 4$),  $\FeAR_{1,1} = Z(5/4) = 1.0$. Similarly for 2, $\FeAR_{2,2}= Z(7/6)=1.0$. Thus, in this case, since both agents have more feasible moves than afforded by other agents choosing MdR, they have the maximum possible FeAR values. 

In instance (b) 1 chooses R1 which leaves 2 with $\nij=5$ feasible moves (S0, R1, R2, R3 and R4) , and 2 chooses R1 which leaves 1 with $\nij=5$ feasible moves (L2, L1, S0, R1 and R2). Thus, $\FeAR_{1,2}= \frac{6-5}{6}=\frac{1}{6}=0.17$ is positive and 1 is being assertive to 2, while, $\FeAR_{2,1} = \frac{4-5}{4}= -0.25$ is negative and 2 is being courteous to 1. This causes a reduction in $\FeAR_{2,2}=5/6= 0.83$ while $\FeAR_{1,1} = Z(5/4) = 1$ remains at the maximum possible value. 

In instance (c) 1 chooses R1 which leaves 2 with $\nij=5$ feasible moves (S0, R1, R2, R3 and R4), and 2 chooses L1 which leaves 1 with $\nij=3$ feasible moves (L2, L1 and S0). Thus, $\FeAR_{1,2}= \frac{6-5}{6}=0.17$ and $\FeAR_{2,1}=\frac{4-3}{4}=0.25$ are both positive which means that both agents are being assertive to each other. This leads to reductions in $\FeAR_{1,1}=5/6=0.83$ and $\FeAR_{2,2}= 3/4 = 0.75$.

The above examples demonstrate how the actions of agents, determine the FeAR values. In the given locations and $\JointMdR = \text{S0-S0}$, the actions L1 of 1 and R1 of 2 are considered courteous from the calculated FeAR values, while the actions R1 of 1 and L1 of 2 are considered assertive from the FeAR values. $\FeAR_{1,2}$ is dependant on the move of 1 and $\FeAR_{2,1}$ is dependant on the move of 2. On the other hand, $\FeAR_{1,1}$ is dependant on the move of 2 and $\FeAR_{2,2}$ is dependant on the move of 1. So, $\FeAR_{i,j}$ where $i \neq j$ capture the assertiveness/courteousness of $i$ towards $j$ , while $\FeAR_{i,i}$ captures the complement of the collective assertiveness/courteousness of all other agents on $i$ . Thus, the chosen moves in combination with context determine the FeAR values and how agents are considered courteous or assertive. 

The examples also showcase how the causal responsibility ascribed to agents can be asymmetric. In (a), both agents are courteous. In (c) both agents are assertive. In (b) one agent is assertive while the other is courteous. It is also worth noting that even in the case where both agents are found to be courteous (a), there is a small asymmetry in the FeAR values ($\FeAR_{1,2}= -0.17$ and $\FeAR_{2,1}=-0.25$). This stems from the asymmetry in the context --- 1 is more restricted than 2 and has fewer feasible moves when considering the MdRs. Thus, the causal influence of agents can be asymmetric and are highly dependent on the context.

\subsection{Two agents on a lane: the effect of agents' Move de Rigueur on FeAR}

\begin{figure*}
	\centering
		\begin{subfigure}{0.4\linewidth}
			\centering
			\caption*{Grid world instance - state and action selection}
		\end{subfigure}
		\hfill
		\begin{subfigure}{0.12\linewidth}
			\centering
			\caption*{$\JointMdR$: S0-S0}
		\end{subfigure}
		\hfill
		\begin{subfigure}{0.12\linewidth}
			\centering
			\caption*{$\JointMdR$: R1-R1}
		\end{subfigure}
		\hfill
		\begin{subfigure}{0.12\linewidth}
			\centering
			\caption*{$\JointMdR$: R2-R2}
		\end{subfigure}
\hrule
\hrule
\vspace{0.1cm}

		\begin{subfigure}{0.4\linewidth}
			\centering
			\includegraphics[width=\linewidth]{Figures/Sim/SingleLane10-2Agents-AgentsAt-3-5_S0-S0_ex_mdr-S0-S0_Instance_11_GW.png}
			\caption{}
			\label{fig:2Agents4MdR-L1-R1}
		\end{subfigure}
		\hfill
		\begin{subfigure}{0.12\linewidth}
			\centering
			\includegraphics[width=\linewidth]{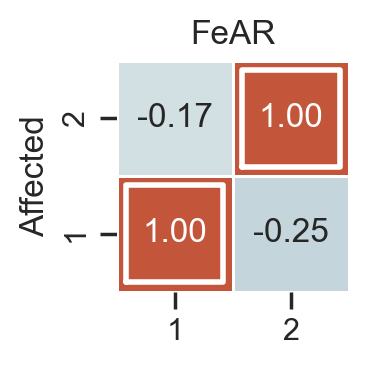}
		\end{subfigure}
		\hfill
		\begin{subfigure}{0.12\linewidth}
			\centering
			\includegraphics[width=\linewidth]{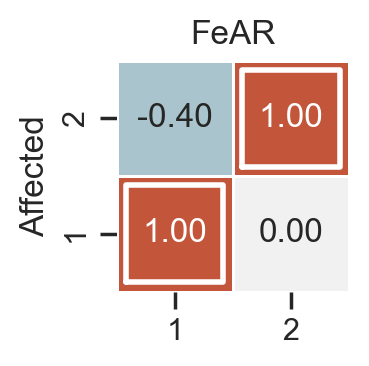}
		\end{subfigure}
		\hfill
		\begin{subfigure}{0.12\linewidth}
			\centering
			\includegraphics[width=\linewidth]{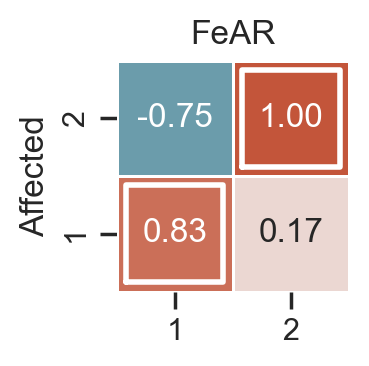}
		\end{subfigure}
\hrule
\vspace{0.01cm}

		\begin{subfigure}{0.4\linewidth}
		\centering
		\includegraphics[width=\linewidth]{Figures/Sim/SingleLane10-2Agents-AgentsAt-3-5_S0-S0_ex_mdr-S0-S0_Instance_20_GW.png}
		\caption{}
		\label{fig:2Agents4MdR-R1-R1}
		\end{subfigure}
		\hfill
		\begin{subfigure}{0.12\linewidth}
			\centering
			\includegraphics[width=\linewidth]{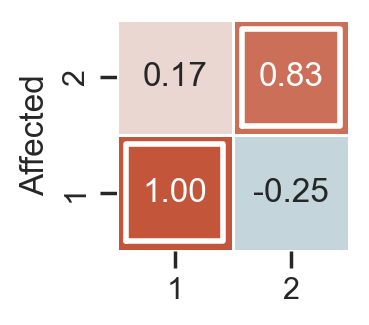}
		\end{subfigure}
		\hfill
		\begin{subfigure}{0.12\linewidth}
			\centering
			\includegraphics[width=\linewidth]{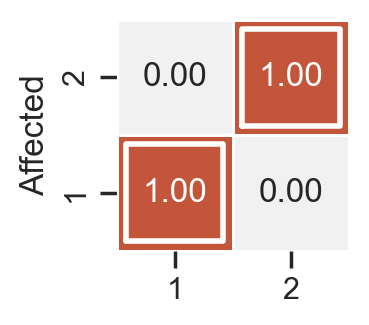}
		\end{subfigure}
		\hfill
		\begin{subfigure}{0.12\linewidth}
			\centering
			\includegraphics[width=\linewidth]{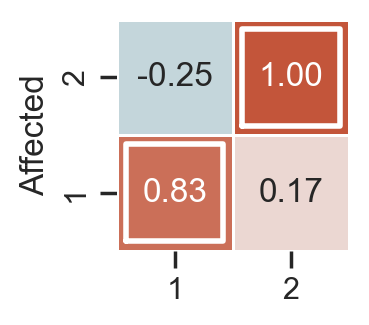}
		\end{subfigure}
\hrule
\vspace{0.01cm}

		\begin{subfigure}{0.4\linewidth}
		\centering
		\includegraphics[width=\linewidth]{Figures/Sim/SingleLane10-2Agents-AgentsAt-3-5_S0-S0_ex_mdr-S0-S0_Instance_19_GW.png}
		\caption{}
		\label{fig:2Agents4MdR-R1-L1}
		\end{subfigure}
		\hfill
		\begin{subfigure}{0.12\linewidth}
			\centering
			\includegraphics[width=\linewidth]{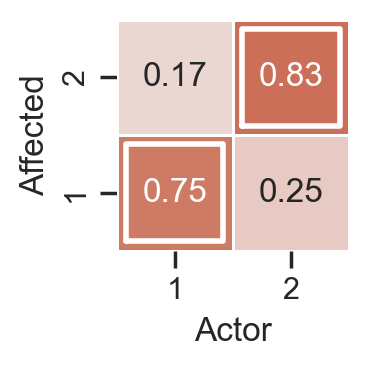}
		\end{subfigure}
		\hfill
		\begin{subfigure}{0.12\linewidth}
			\centering
			\includegraphics[width=\linewidth]{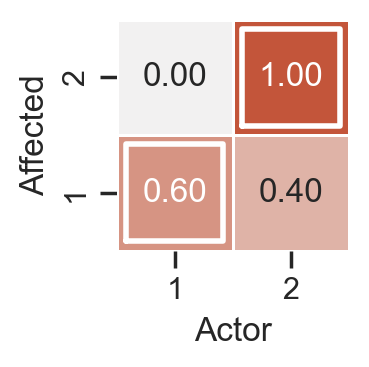}
		\end{subfigure}
		\hfill
		\begin{subfigure}{0.12\linewidth}
			\centering
			\includegraphics[width=\linewidth]{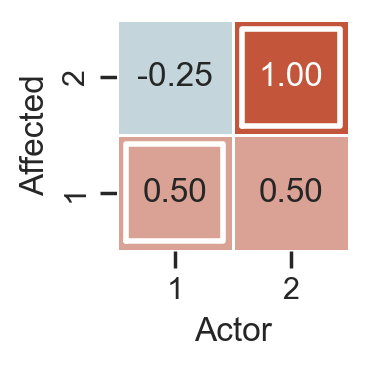}
		\end{subfigure}
\hrule
\vspace{0.01cm}

	\caption[2 agents on a lane - different MdRs]{\textbf{Case Study 2} explores the effect of MdR on the FeAR metric. We consider the same instances as in Case Study 1 (\cref{fig:2Agents}). Here three different MdRs are considered --- $\JointMdR=\text{S0-S0}$, $\JointMdR=\text{R1-R1}$ and $\JointMdR=\text{R2-R2}$. $\FeAR_{i,j}$ for $i \neq j$, represents the Feasible Action-Space Reduction of $j$ by the action of $j$. $\FeAR_{i,i}$ represent Feasible Action-Space Remaining for agent $i$ left by the actions of all other agents. For $i \neq j$, if $\nij> \nijMdR$, then $\FeAR_{i,j} < 0$ and vice versa. Thus changes in $\nijMdR$ affect the FeAR values.}
	\label{fig:2Agents4MdR}
\end{figure*}

In the first case study, we considered only one MdR, $\JointMdR = \text{S0-S0}$. In this case study, we explore how changing the MdRs can affect the FeAR values, again for the same instances considered before. $\JointMdR = \text{R1-R1}$ considers the norm where both agents are supposed to move one step to the right --- like on a lane where cars are not supposed to stop in the middle of the lane. $\JointMdR = \text{R2-R2}$ considers a norm where the agents are supposed to keep moving --- now at a higher speed than when $\JointMdR=\text{R1-R1}$. For ease of reference, we are repeating the FeAR values for $\JointMdR = \text{S0-S0}$ here too (See \cref{fig:2Agents4MdR}). For instances (a), (b) and (c) the feasible actions available to agents 1 and 2 for each case remain the same as they were for the previous case study.

For $\JointMdR=\text{S0-S0}$, 1 has $\nijMdR= 4$ feasible moves (L2, L1, S0 and R1) when 2 chooses the MdR, and 2 has $\nijMdR= 6$ feasible moves (L1, S0, R1, R2, R3 and R4) when 1 chooses the MdR. For $\JointMdR=\text{R1-R1}$, 1 has $\nijMdR= 5$ feasible moves (L2, L1, S0, R1 and R2) when 2 chooses the MdR (R1), and 2 has $\nijMdR= 5$ feasible moves (S0, R1, R2, R3 and R4) when 1 chooses the MdR (R1). For $\JointMdR=\text{R2-R2}$, 1 has $\nijMdR= 6$ feasible moves (L2, L1, S0, R1, R2 and R3) when 2 chooses the MdR (R2), and 2 has $\nijMdR= 4$ feasible moves (R1, R2, R3 and R4) when 1 chooses the MdR (R2).

The move L1 of 1 (instance (a)) offers more feasible actions to 2 than all the MdRs considered (S0, R1 and R2). Thus, when 1 chooses L1, $\FeAR_{1,2}$ for all the MdRs are negative and indicate that 1 is being courteous. It can also be noted that the move L1 of 1 is considered more courteous under $\JointMdR=\text{R2-R2}$ than under $\JointMdR=\text{S0-S0}$  (and $\JointMdR=\text{R1-R1}$ being an intermediate condition).

Similarly, the move L1 of 2 (instance c)) offers fewer feasible actions to 1 than all the MdRs considered (S0, R1 and R2). Thus when 2 chooses L1, $\FeAR_{2,1}$ for all the MdRs are positive and indicate that 2 is being assertive. Also, as we go from $\JointMdR = \text{S0-S0}$ to $\JointMdR = \text{R2-R2}$, the 2 choosing L1 is considered increasingly assertive ($\FeAR_{2,1}$ for $\JointMdR=\text{S0-S0}$ $<$ $\FeAR_{2,1}$ for $\JointMdR=\text{R1=R1}$ $<$$\FeAR_{2,1}$ for $\JointMdR=\text{R2-R2}$). 

As we go from $\JointMdR = \text{S0-S0}$ to $\JointMdR = \text{R2-R2}$, the move R1 of 1 (instances (b and c)) goes from being considered assertive ($\FeAR_{1,2}>0$) to being considered courteous($\FeAR_{1,2}<0$), while, the move R1 of 2 (instances(a and(b)) goes from being considered courteous  ($\FeAR_{2,1}<0$) to being considered assertive ($\FeAR_{1,2}>0$). These examples illustrate how the complex interplay of actions, contexts and norms (represented by the MdRs) can be represented by the FeAR metric.

\subsection{Three agents at an Intersection: the effect of other agents' actions on FeAR}
\begin{figure}[t]
	\centering
	\begin{subfigure}{0.5\linewidth}
		\centering
		\caption*{Grid world instance}
	\end{subfigure}
	\begin{subfigure}{0.34\linewidth}
		\centering
		\caption*{$\JointMdR$: S0-S0-S0}
	\end{subfigure}

\hrule \hrule \vspace{0.1cm}

	\begin{subfigure}{0.5\linewidth}
		\centering
		\includegraphics[width=\linewidth]{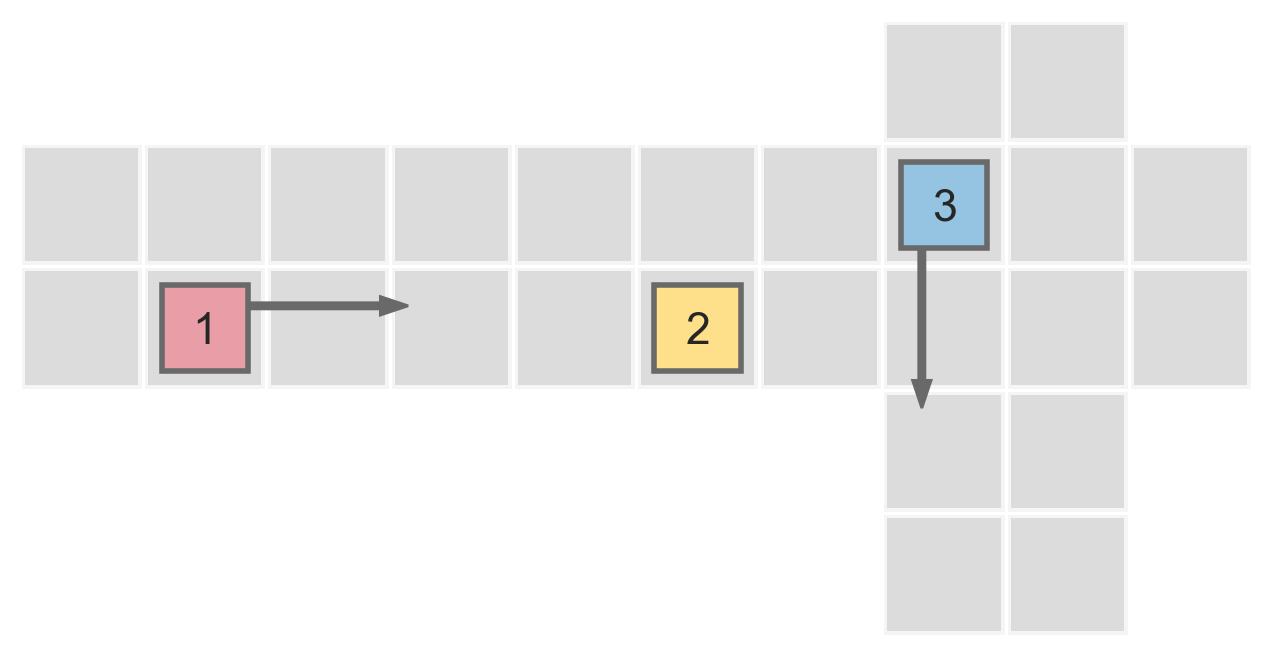}
		\caption{No collisions}
		\label{fig:Intersection-R2-S0-D2}
	\end{subfigure}
	\begin{subfigure}{0.34\linewidth}
		\centering
		\includegraphics[width=\linewidth]{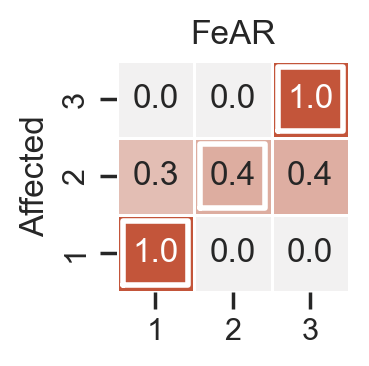}
	\end{subfigure}
\hrule \vspace{0.1cm}

	\centering
	\begin{subfigure}{0.5\linewidth}
		\centering
		\includegraphics[width=\linewidth]{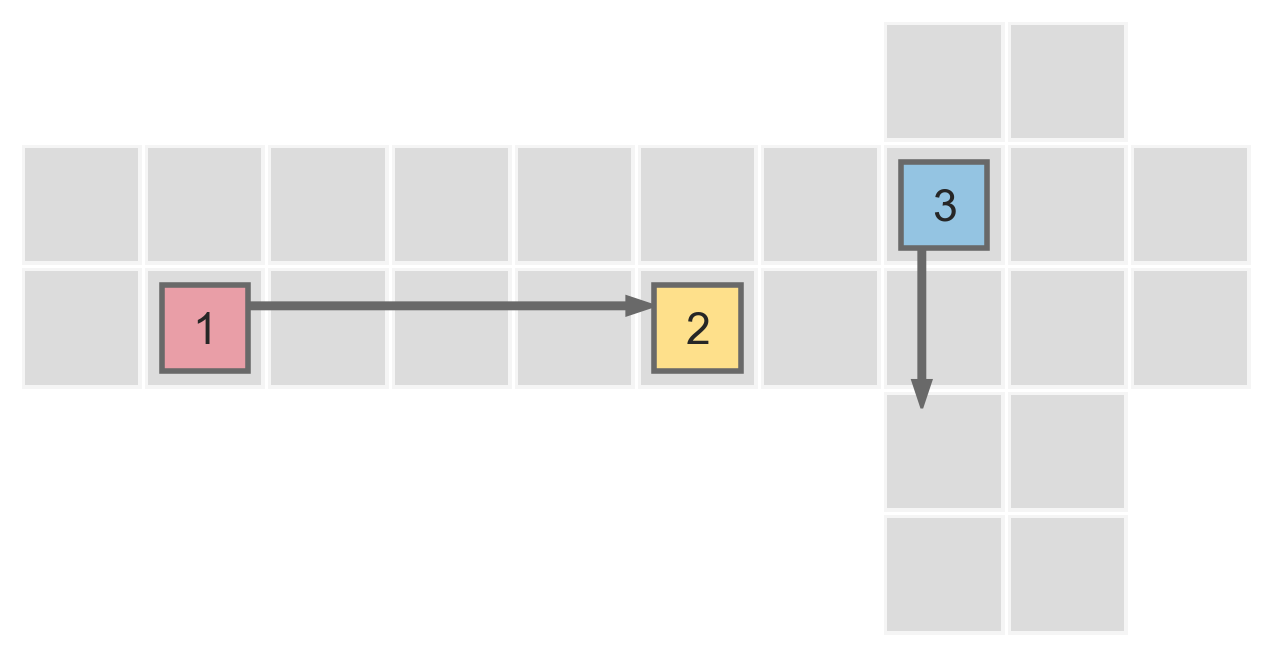}
		\caption{\textcolor{BrickRed}{Collision} between 1 and 2}
		\label{fig:Intersection-R4-S0-D2}
	\end{subfigure}
	\begin{subfigure}{0.34\linewidth}
		\centering
		\includegraphics[width=\linewidth]{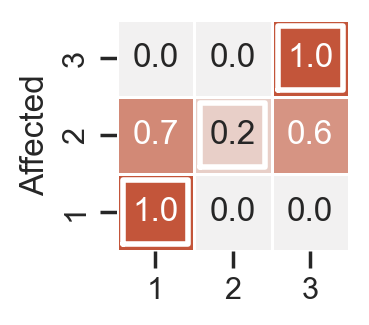}
	\end{subfigure}
\hrule \vspace{0.1cm}

	\centering
	\begin{subfigure}{0.5\linewidth}
		\centering
		\includegraphics[width=\linewidth]{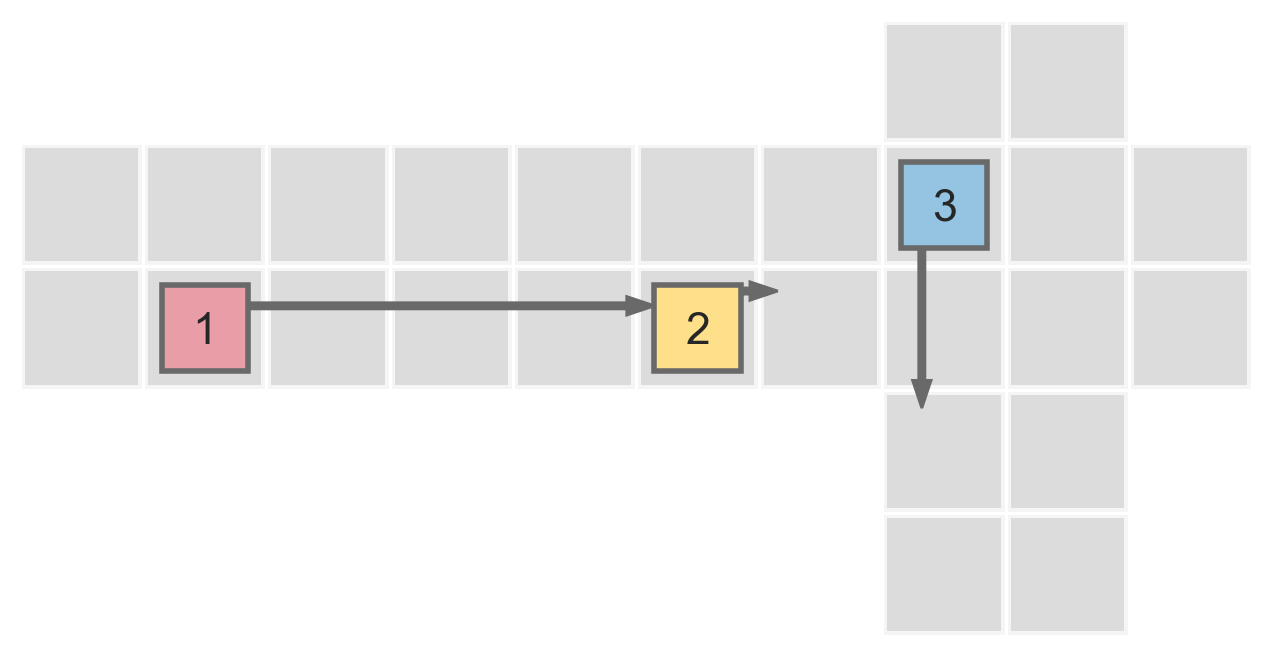}
		\caption{No collisions}
		\label{fig:Intersection-R2-R1-D2}
	\end{subfigure}
	\begin{subfigure}{0.34\linewidth}
		\centering
		\includegraphics[width=\linewidth]{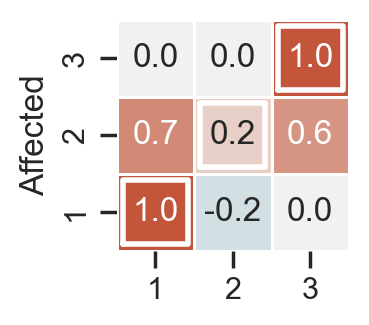}
	\end{subfigure}
\hrule \vspace{0.1cm}

	\centering
	\begin{subfigure}{0.5\linewidth}
		\centering
		\includegraphics[width=\linewidth]{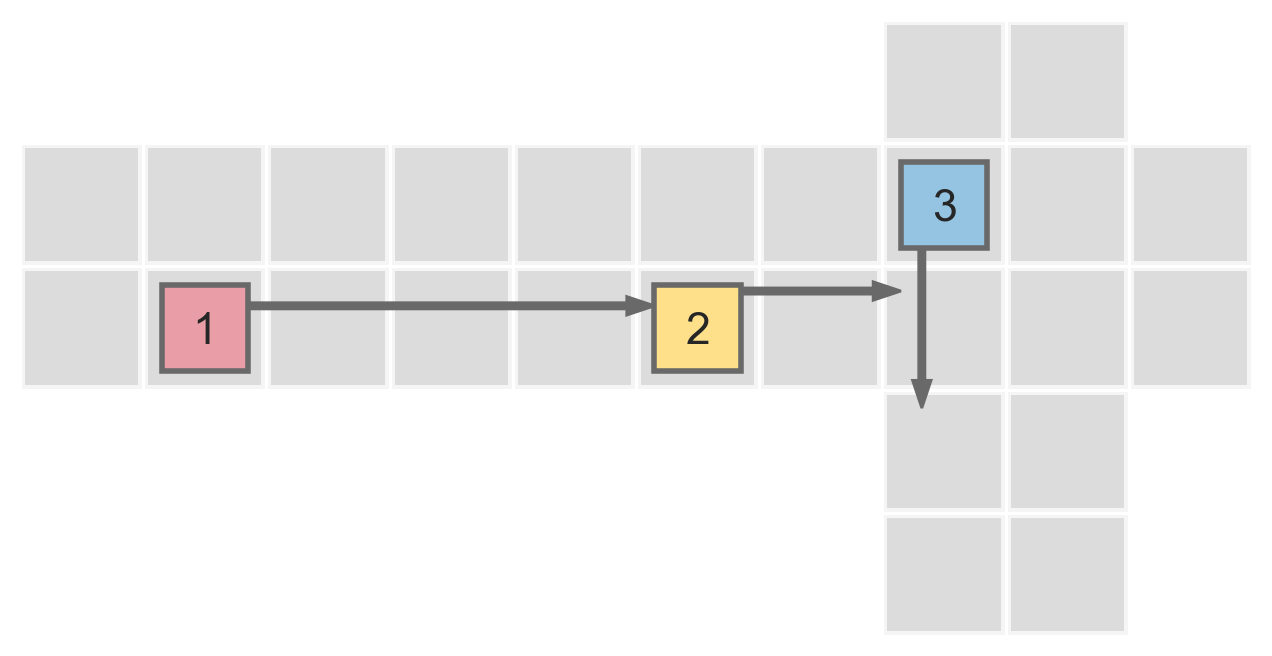}
		\caption{\textcolor{BrickRed}{Collision} between 2 and 3}
		\label{fig:Intersection-R4-R2-D2}
	\end{subfigure}
	\begin{subfigure}{0.34\linewidth}
		\centering
		\includegraphics[width=\linewidth]{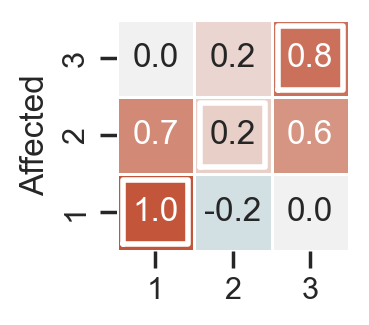}
	\end{subfigure}
\hrule \vspace{0.1cm}

	\centering
	\begin{subfigure}{0.5\linewidth}
		\centering
		\includegraphics[width=\linewidth]{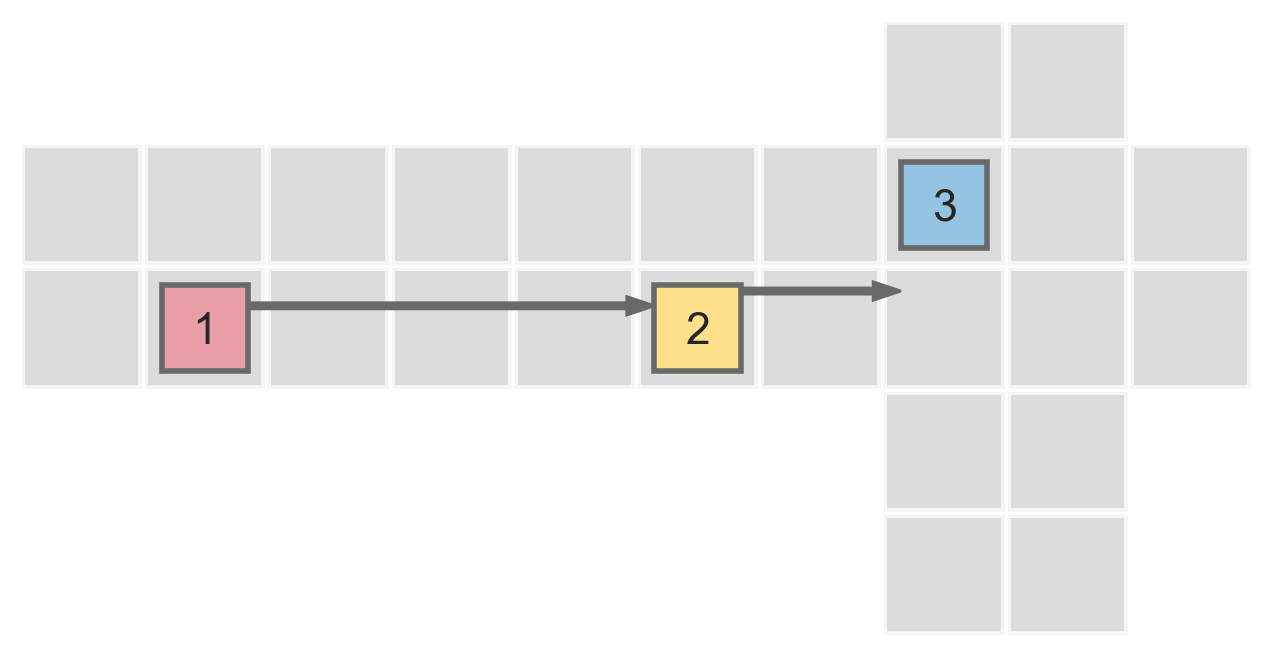}
		\caption{No collisions}
		\label{fig:Intersection-R4-R2-S0}
	\end{subfigure}
	\begin{subfigure}{0.34\linewidth}
		\centering
		\includegraphics[width=\linewidth]{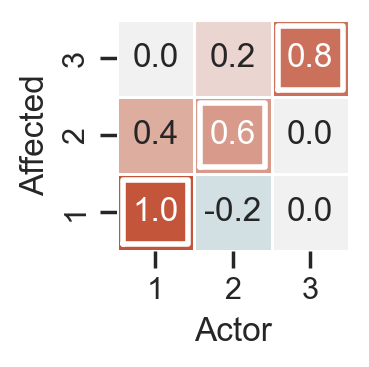}
	\end{subfigure}
\hrule \vspace{0.1cm}

\caption[Intersection with 3 Agents]{\textbf{Case Study 3} explores the FeAR metric in the context of multi-agent interactions. We consider the interactions between three agents at an intersection. Agent 2 is caught between Agents 1 and 3. Instance (c) is the central condition where no collisions happen. But, if agent 2 slightly changes its move, it can collide with agent 1 (instance (b)) or with agent 3 (instance (d)). The collisions in instances (b) and (d) could also be prevented by the actions of agent 1 (instance (a)) or agent 3 (instance (d)) respectively. Thus, in these interactions $\FeAR_{1,2}$ and $\FeAR_{3,2}$ are dependent on the actions chosen by third parties too.}
\label{fig:Intersection3Agents}
\end{figure}

 This case study illustrates the use of the FeAR metric in interactions of multiple agents. We consider interactions of three agents at an intersection (\cref{fig:Intersection3Agents}). Agents 1 and 2 are approaching the intersection from the left and agent 3 is approaching it from above. In instance (a), there are no collisions and agent 3 safely passes the intersection. If agent 1 decides to move faster (R4), it causes agent 1 to rear-end agent 2 (instance (b)). This rear-ending can be prevented if agent 2 moves forward (R1) as seen in instance (c). But, if agent 2 moves forward faster (R2), it causes a T-bone collision with agent 3 (instance (d)). This T-bone collision could have been prevented if agent 3 had chosen to stay (S0) as in instance (e). Thus, we explore a series of interactions where agent 2 is caught between agents 1 and 3. 

 As we move from instances (a) to (b), the action of agent 1 becomes more restrictive to agent 2 and $\FeAR_{1,2}$ increases from $0.3$ to $0.7$. This value of $\FeAR_{1,2}$  remains unchanged for instances (b), (c) and (d) --- irrespective of whether agent 2 collides with agent 1 or 3. In instance (e), when agent 3 chooses (S0), it increases the feasible action space of 2. This leads to $\FeAR_{1,2}$ decreasing to $0.4$ even though the action of agent 1 does not change from instances (b), (c) and (d). Thus, $\FeAR_{1,2}$ is dependent on the actions of both agents 1 and 3. Similarly, we can see the influence of agent 1's action on $\FeAR_{3,2}$ as we go through instances (e), (d), (c), (b) and (a) ($\FeAR_{3,2} = 0.0$ for (e), $\FeAR_{3,2}=0.6$ for (d), (c) and (b), and $\FeAR_{3,2} = 0.4$ for (a)). 

 Now, let us consider the effect of the actions of agent 2 on agents 1 and 3. In instances (a) and (b), agent 2 is choosing S0 which is the MdR. Hence, $\FeAR_{2,1}=0$ and $\FeAR_{2,3}=0$. When agent 2 chooses R1 in instance (c), it increases the feasible action space of 1 and thus $\FeAR_{2,1}$ decreases to $-0.2$. Since choosing R1 does not put agent 2 in the path of agent 3, the feasible action space of agent 3 remains unchanged and $\FeAR_{2,3}=0$. But, when agent 2 chooses R2 (instances (d) and (e)), it moves into the path of agent 3 restricting the feasible action space of agent 3. Thus, $\FeAR_{2,3}$ increases to $0.2$. Thus in instances (d) and (e), agent 2 is acting assertively towards agent 3 ($\FeAR_{2,3}>0$) and courteously towards agent 1 ($\FeAR_{2,1}$)

 In our analysis, we only considered one MdR $\JointMdR = \text{S0-S0-S0}$ which gives equal priority at all. In real traffic, certain lanes are given priority over others at intersections. These changes in priorities could be captured using MdRs like $\JointMdR = \text{S0-S0-D2}$ (which prioritises the vertical lane) or $\JointMdR = \text{R2-R2-S0}$ (which prioritises the horizontal lane). Analyses of these scenarios can be found in the supplementary materials.
 
 Overall, this cases study illustrates how the complexities of multi-agent interactions are captured in the FeAR metric. We also show how agents caught in the middle can be assertive to some agents while being courteous to other agents.

\subsection{Four agents on a lane: illustrating complex interactions}

\begin{figure}[t]
    \centering
    \begin{subfigure}{0.9\linewidth}
       \captionsetup{justification=centering}
    \caption*{Instances with extreme values of $\left(\sum_{i}^{k} \sum_{\substack{j=1\\j \neq i}}^k (\FeAR_{i,j})^2\right)$}
    \end{subfigure}
    \hrule \hrule \vspace{0.1cm}
    \begin{subfigure}[c]{0.64\linewidth}
        \centering
        \includegraphics[width=\linewidth]{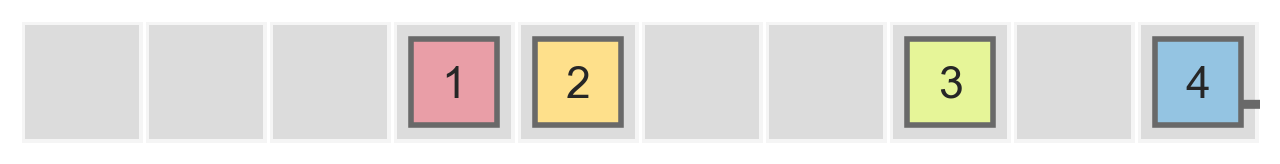}
    \caption{Instance with smallest sum of non-diagonal FeAR values}
    \label{fig:4Agents-MinFeAR}
    \end{subfigure}
    \begin{subfigure}[c]{0.35\linewidth}
        \centering
        \includegraphics[width=\linewidth]{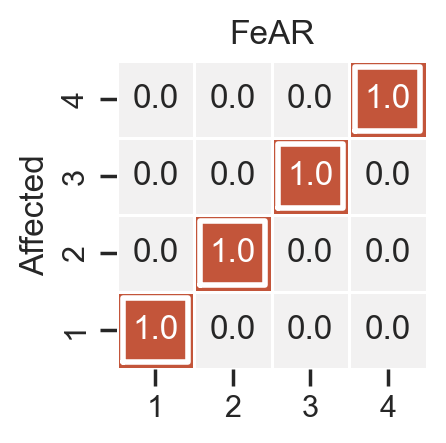}
    \end{subfigure}

    \begin{subfigure}[c]{0.64\linewidth}
        \centering
        \includegraphics[width=\linewidth]{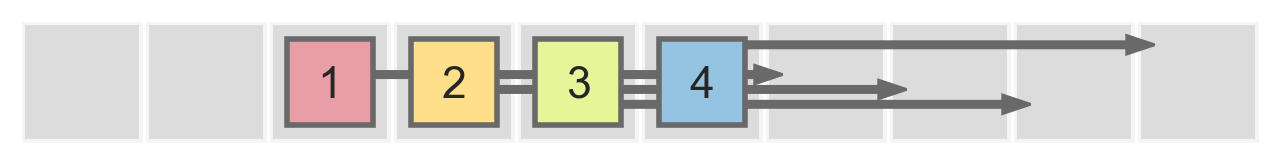}
    \caption{Instance with biggest sum of non-diagonal FeAR values}
    \label{fig:4Agents-MaxFeAR}
    \end{subfigure}
    \begin{subfigure}[c]{0.35\linewidth}
        \centering
        \includegraphics[width=\linewidth]{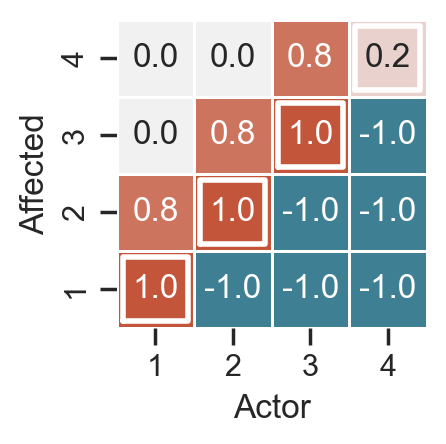}
    \end{subfigure}
    
    \hrule
    \caption{\textbf{Case Study 4} illustrates the application of the metric in scenarios with potentially complex interactions of multiple agents. These two instances were selected from 10,000 instances based on the FeAR values. In \cref{fig:4Agents-MinFeAR} --- the case with minimum sum of squares of FeAR --- all agents choose MdR (or the action equivalent to MdR due to hitting a border). In \cref{fig:4Agents-MaxFeAR} --- the case with maximum sum of square of FeAR --- agents are acting assertively to the agents ahead of them, while being courteous to the agents behind them. All of them deviate from the MdR and together they avoid collisions.}
\end{figure}

To highlight the insights provided by the FeAR metric in more complex scenarios, we analysed the interactions of 4 agents on a single lane. To do this, we randomly generated 10,000 instances that all involved 4 agents on a lane 10 units long. We randomized the locations of the agents as well as their action choices (S0, R1, R2, R3 and R4). This set of actions was selected to mimic how agents would behave on a one-way lane. For each agent, the location on the lane and the action were chosen at random with equal likelihood.  

We expected these randomly generated instances to vary substantially in complexity of the interaction. Here we investigated two extreme cases: an instance in which the agents acted most independently from each other, and an instance with a high level of interconnectedness between the agents. To do so, among the 10,000 generated instances we selected the ones with the minimum and maximum sum of squares of the $\FeAR_{i,j}$ for $i \neq j$ (\cref{fig:4Agents-MinFeAR} and \cref{fig:4Agents-MaxFeAR}, respectively).

In the minimum sum-of-squares case \cref{fig:4Agents-MinFeAR}, agents 1,2 and 3 choose S0; agent 4's move to the right hits the boundary. Thus all agents would effectively be doing S0. Which means that effectively all agents are following the MdR and hence FeAR values are zero. It must be noted that this is not the only case where the sum of squares of FeAR is zero (the others are not shown here).

In the maximum sum-of-squares case \cref{fig:4Agents-MaxFeAR}, there are no collisions, but all the agents are deviating from the MdR (S0). If agents 2, 3 or 4 choose the MdR, they would cause a collision for all the agents behind them. Thus their current action increases the feasible action space of all agents behind them --- indicated by the negative FeAR values below the diagonal. At the same time, since agents 1,2 and 3 choose R4, they severely restrict the feasible action of the agent in front of them, compared to when they choose the MdR.

These two cases demonstrate that in multi-agent interactions, the 
sum of non-diagonal FeAR values captures the intuitions about the independence/interconnectedness of the agents' actions. In a similar vein, large positive (non-diagonal) FeAR values across multiple agents would flag cases where agents are being very assertive. Conversely, instances where agents are being very courteous would be marked by consistently negative values of FeAR (see the supplementary materials for details). 

Overall, this case study illustrates how the FeAR metric can be analysed in more complex scenarios in which the number of possible combinations of agent positions and actions is very large (more than 3 million in this case). To this end, comparisons between different outcomes of the interaction (e.g. different actions of the agents for a given set of positions) can be facilitated by aggregate FeAR values across all the agents.

\section{Discussion and Conclusion}
\label{Sec:Discussion}

Our Feasible Action-space Reduction metric can be used as an indicator for causal responsibility in spatial interactions where each agent is physically present in the scene and, therefore, has to take at least one action. The FeAR metric captures the effects of actions of agents, norms (represented as \textit{Moves de Rigueur} (MdR)), and the physical layout of the grid world. The case studies also illustrate how the responsibility ascription among agents can be asymmetric ($\FeAR_{i,j} \neq \FeAR_{j,i}$). We also illustrate how agents can be assertive to some agents ($\FeAR_{i,j}>0$), while being courteous to others ($\FeAR_{i,j}<0$). A positive FeAR value does not necessarily mean that a collision happens, but if a collision does happen, it indicates the share of responsibility of a given agent. For example, with three agents interacting at an intersection \cref{fig:Intersection3Agents}, when two agents (1 and 3) collectively constrain an agent (2), irrespective of whether it collides with 1 or 3, both agents share some responsibility for the state of 2. In real-world traffic interactions, when a car swerves to avoid hitting another car but ends up hitting a third car, it can be unclear who is responsible for the interaction. The FeAR metric along with a suitably chosen set of MdRs would be able to help guide the debate on the distribution of responsibility in such controversial cases.

We used Moves de Rigueur to represent norms. MdRs are actions that are expected of or prescribed to agents in an interaction. Ideally, theses actions should be aligned with human behaviour and cognitive capacity. For example, consider the case where humans are used/trained to stop at an intersection. In this case, prescribing an MdR to keep driving through an intersection would be inappropriate. Similarly, automated vehicles require fallback actions and minimal risk manoeuvres to ensure safety in case of emergencies. Well-defined MdRs aligned with human behaviour could provide good fallback actions for automated-vehicle algorithms. 

We proposed Feasible Action-Space Reduction as a metric for causal responsibility among agents in a spatial interaction. While other models of responsibility \cite{bartlingShiftingBlameDelegation2012,chocklerResponsibilityBlameStructuralModel2004,douerResponsibilityQuantificationModel2020,englTheoryCausalResponsibility2018,triantafyllouActualCausalityResponsibility2022,yazdanpanahDistantGroupResponsibility2016} ascribe responsibility for events, in spatial interactions, the FeAR metric assigns causal responsibility for the state and safety of agents. Causal responsibility has traditionally been evaluated based on  probabilistic reasoning over sequences of events. Our formulation does not consider the probabilities of actions --- we consider that all actions of an agent are equally likely. Our metric only considers an instant of the interaction and does not consider the history of events like other approaches. Because of the different nature of these metrics, no straightforward comparison can be made between FeAR and other models of causal responsibility. Developing methods to meaningfully compare disparate ways of quantifying responsibility is an important avenue for future work. 

Existing literature on spatial navigation has several theories and models that try to address various aspects of how pedestrians and drivers engage in spatial interactions \cite{helbingSocialForceModel1995, kolekarHumanlikeDrivingBehaviour2020, markkulaExplainingHumanInteractions2023, pekkanenVariableDriftDiffusionModels2022, schwartingSocialBehaviorAutonomous2019, siebingaModellingCommunicationenabledTraffic2023}. Such models, typically incorporate several assumptions to explain and predict human behaviour. In contrast, the goal of this paper is not to describe human behaviour but to develop a model-agnostic framework for quantifying causal responsibility. However, we do believe that such models would be crucial for anticipating the behaviour of other agents when using FeAR for estimating responsibility of future actions (ex-ante responsibility).

Methods like Responsibility Sensitive Safety~\cite{shalev-shwartzFormalModelSafe2018} prescribe "proper responses" for all agents and ascribe greater responsibility to agents that deviate from the "proper response". The MdR approach offers more freedom to the agents --- agents do not have to follow the MdR strictly and additionally have the possibility of being more courteous than the MdR. Furthermore, designing consistent MdRs can help humans and automation to have mutually compatible representations, which can help reduce the responsibility gaps in human-AI interactions~\cite{cavalcantesiebertMeaningfulHumanControl2022, santonidesioMeaningfulHumanControl2018}. 

\noindent
\textbf{Limitations and future work.}
FeAR does not look at past interactions, states or beliefs of the agents. To get a more complete picture of causal responsibility as well as capture moral/role/capacity responsibility~\cite{vincentStructuredTaxonomyResponsibility2011}, more factors should be considered. The current implementation of FeAR only works for discrete and finite action and state spaces. In real-life interactions, an agent might have an uncountable number of actions to choose from. Our goal was to provide the first steps and the foundation towards measuring causal responsibility in spatial settings, and we felt that adding more complexity would have made the problem intractable. By making these assumptions, we were able to derive a metric that will then serve as a basis for investigating more complex settings, including continuous time and action space. Also, the current formulation of the metric scales exponentially with the number of agents and thus is not practical for real-time applications. A possible improvement point for future works would be to consider the calculated FeAR values as ground truth to train a supervised learning method to approximate the FeAR in a given instance.

Defining MdRs for all possible scenarios is a non-trivial task; especially if done manually. Establishing good norms for driving (like, ‘always maintain safe time headway’) would limit occurrence of scenarios where defining MdRs is challenging. Nevertheless, there might be situations where defining MdRs ahead of time might not be feasible, due to unknown circumstances or unpredictable behaviour of agents. Methods like Norm Synthesis \cite{moralesAutomatedSynthesisNormative2013,moralesSynthesisingLiberalNormative2015} or Responsibility-Sensitive Safety \cite{shalev-shwartzFormalModelSafe2018,shalev-shwartzVisionZeroProvable2019} could be useful to automate the process of defining conflict-free MdRs. Making the MdR method applicable in practice is a challenging issue that future work should address.

We opted to present the metric in an intuitive way and use case studies to evaluate such intuitions and how it could be applied for understanding causal responsibility in multi-agent spatial interactions. In future works, especially when dealing with continuous domains, we believe that a formal evaluation might be necessary to achieve necessary insights on the robustness of the metrics and its impact on safety

\textbf{Conclusions.}
We proposed Feasible Action-space Reduction (FeAR) as a metric for quantifying the causal responsibility a given agent's actions have on the state and safety of other agents in an interaction. To represent norms, we introduced \textit{Moves de Rigueur} (MdR) ---  consistent sets of default actions that are prescribed to each agent in an interaction. We used case studies to illustrate how the metric captures the effects of agents' actions, norms and multi-agent interactions. We also demonstrated an application of the metric to flag instances with high levels of interconnectedness/independence among the actions chosen by agents. 

Ensuring safety and implementing meaningful human control in automated driving requires a common interdisciplinary framework that engages conversations between road users, vehicle manufacturers, traffic designers, regulatory agencies and policy makers \cite{calvertDesigningAutomatedVehicle2023,papadimitriouCommonEthicalSafe2022}. The FeAR metric and the concept of MdR is a step towards this interdisciplinary framework.

\ack This work is supported by the TU Delft AI Labs programme.

\bibliography{EmergenceAndResponsibility}

\end{document}

%% file: agdefs.tex

\newif\ifOnlySummary
\OnlySummarytrue


\newcommand{\StateSpace}{\mathcal{S}}
\newcommand{\Agents}{\mathcal{K}}
\newcommand{\JointActionSpace}{\mathcal{A}}
\newcommand{\ActionSpace}[1]{\mathcal{A}_{#1}}
\newcommand{\JointMdR}{\mu}

\newcommand{\aState}{s}
\newcommand{\jointAction}{A}
\newcommand{\action}[1]{a_{#1}}
\newcommand{\mdr}[1]{\mu_{#1}}
\newcommand{\validCount}[3]{n \bigl(#1, #2, #3 \bigr)}

\newcommand{\nij}{\validCount{\aState}{\jointAction}{j}}
\newcommand{\nijMdR}{\validCount{\aState}{\left[ \jointAction_{i} \leftarrow \mdr{i} \right]}{j}}
\newcommand{\nii}{\validCount{\aState}{\jointAction}{i}}
\newcommand{\niiMdR}{\validCount{\aState}{\left[ \jointAction_{\neg i} \leftarrow \mdr{\neg i} \right]}{i}}

\newcommand{\FeAR}{\mathrm{FeAR}}
\newcommand{\clip}[1]{Z \Biggl(#1\Biggr)}

